\documentclass[preprint,12pt]{elsarticle}

\usepackage[T1]{fontenc}
\usepackage[utf8]{inputenc}
\usepackage{lmodern}
\usepackage{amsmath,amssymb,amsfonts}
\usepackage{graphicx}
\usepackage{booktabs}
\usepackage{hyperref}
\usepackage{xcolor}
\usepackage{float}
\usepackage{algorithm}
\usepackage{algpseudocode}

\biboptions{sort&compress}

\journal{Computers \& Fluids}

\makeatletter
\def\ps@pprintTitle{%
  \let\@oddhead\@empty
  \let\@evenhead\@empty
  \def\@oddfoot{\footnotesize\itshape Preprint}%
  \let\@evenfoot\@oddfoot
}
\makeatother

\begin{document}

\begin{frontmatter}

\title{Physics-Constrained Neural Closure for Lattice Boltzmann Large-Eddy Simulation}

\author[label1]{Muhammad Idrees Khan\corref{cor1}}
\ead{muhammadidrees.khan@students.uniroma2.eu}

\author[label2,label3]{Sauro Succi}

\author[label4]{Hua-Dong Yao}

\cortext[cor1]{Corresponding author.}

\affiliation[label1]{organization={Department of Enterprise Engineering ``Mario Lucertini'', University of Rome ``Tor Vergata''},
            addressline={Via del Politecnico 1},
            city={Rome},
            postcode={00133},
            country={Italy}}

\affiliation[label2]{organization={Italian Institute of Technology},
            addressline={Piazzale Aldo Moro 1},
            city={Rome},
            postcode={00185},
            country={Italy}}

\affiliation[label3]{organization={Department of Physics, Harvard University},
            addressline={33 Oxford Street},
            city={Cambridge},
            postcode={02138},
            state={Massachusetts},
            country={USA}}

\affiliation[label4]{organization={Department of Mechanics and Maritime Sciences, Chalmers University of Technology},
            addressline={},
            city={Gothenburg},
            postcode={41296},
            country={Sweden}}

\author[label1,label3]{Giacomo Falcucci}

\begin{abstract}
We present a physics-constrained, data-driven subgrid-scale (SGS) stress closure for large-eddy simulation (LES) in the lattice Boltzmann method (LBM). Trained on filtered--downsampled (FD) data from LBM direct numerical simulation (DNS) of forced homogeneous isotropic turbulence (FHIT) spanning multiple filter widths, a compact neural network maps nine macroscopic derivative inputs---six strain-rate and three vorticity components---to the six independent components of the SGS stress tensor; a deviatoric projection is applied post-inference to obtain the traceless stress used in the solver. Training combines a stress data loss with physics terms for SGS energy-transfer ($\Pi$) matching, rotational equivariance under cube rotations, and compatibility of the implied SGS forcing with the divergence-based coupling.

The predicted stress is coupled to the solver through a split strategy: a dissipative, strain-aligned contribution is represented through an effective-viscosity projection, while the remaining anisotropic residual is applied through a forcing term. This construction is intended to retain both backscatter (via the effective viscosity) and non-dissipative anisotropic effects (via the residual forcing), while remaining compatible with LBM deployment. In the cases considered here, a priori results show good agreement with FD references across stress components and SGS-transfer statistics, and a posteriori rollouts improve several energetic and statistical measures relative to static and dynamic Smagorinsky baselines. A preliminary transfer test in turbulent channel flow is also reported without retraining. Finally, we demonstrate production deployment via ONNX Runtime, with throughput comparable to a dynamic Smagorinsky baseline in the tested configuration.
\end{abstract}

\begin{highlights}
\item Data-driven ML-LES closure with physics constraints and compact neural architecture
\item 9 features: strain-rate (6) and vorticity (3)
\item Physics-constrained training with $\Pi$-matching, rotational equivariance, and divergence-free SGS-force penalty
\item Hybrid coupling: dissipative part via effective viscosity, anisotropic residual via forcing; captures backscatter
\item ONNX Runtime deployment in Fortran/OpenACC with batched GPU inference
\item A priori and a posteriori validation vs classical SGS models; preliminary transfer test in channel flow without retraining
\end{highlights}


\end{frontmatter}

\section{Introduction}

Large-eddy simulation (LES) is widely used to study turbulent flows at Reynolds numbers for which direct numerical simulation (DNS) remains prohibitively expensive. In LES, the governing equations are spatially filtered and the influence of unresolved scales must be represented through the subgrid--scale (SGS) stress. Classical SGS closures remain attractive because of their robustness and low cost, but they often trade structural fidelity for stability, which limits their ability to represent stress anisotropy and energy backscatter.

Recent data-driven turbulence modeling spans anisotropic Reynolds-stress modeling, supervised SGS closure learning, and reinforcement-learning-based closure discovery. Tensor-valued learning of Reynolds stresses has been explored, for example, in neural-network models for anisotropic Reynolds stresses in turbulent channel flow~\cite{fang2020neural}. Supervised learning of LES closure terms and SGS stresses has been studied extensively in continuum Navier--Stokes LES~\cite{beck2019deep}, while reinforcement-learning approaches have been used to discover SGS closures in LES~\cite{novati2021automating} and wall models in wall-modeled LES (WMLES)~\cite{bae2022scientific}. Related data-driven closure discovery has also recently been explored in coarse-grained lattice Boltzmann method (LBM) settings~\cite{fischer2025optimal}. These developments motivate the present work, but most of them have been formulated in continuum Navier--Stokes solvers rather than in LBM frameworks.

Explicit SGS-stress learning has been studied mainly in continuum Navier--Stokes LES; its use in LBM LES remains much less explored. A key advantage of LBM for LES is that stress and pressure are locally available in space and time from the distribution functions, without requiring global pressure solves; this local availability motivates strong interest in LES/LBM. Whether LBM confers additional computational advantages in training or extends to regimes where continuum solvers fail are open questions beyond the scope of this work. We focus instead on whether an explicit-stress ML closure can be coupled to LBM in a stable and practical way, and what accuracy and runtime it achieves relative to conventional viscosity-only baselines. In the cases examined here, the coupled closure improves stress and energy-transfer statistics and captures backscatter that Smagorinsky suppresses, while achieving throughput comparable to a dynamic Smagorinsky baseline~(\ref{app:implementation}, Table~\ref{tab:runtime_mlups}).

These considerations lead to three questions that guide the paper:
\begin{enumerate}
    \item Can a compact ML model predict the full SGS stress tensor accurately enough to be useful in closed-loop LBM-LES?
    \item Can that predicted stress be coupled back to LBM in a stable and efficient way, despite the kinetic form of the solver?
    \item In the test cases considered here, does such an explicit-stress coupling provide advantages over standard viscosity-only LBM closures, particularly for anisotropy and backscatter?
\end{enumerate}
A broader motivation is whether ML can learn a closure more realistic, general, or accurate than the extended kinetic relaxation-time formulation of Chen \textit{et al.}~\cite{chen2003extended}.

To address these questions, we develop an explicit-stress ML closure tailored to LBM. Training data are generated from LBM DNS of forced homogeneous isotropic turbulence (FHIT) and processed into filtered--downsampled (FD) fields at multiple filter widths to form paired datasets of resolved fields and exact SGS stresses. The neural network is trained on macroscopic derivative features (strain-rate and vorticity components) and predicts the six independent components of the SGS stress tensor; a deviatoric projection is applied post-inference before coupling to the solver. The predicted stress is then introduced through a hybrid scheme: the dissipative, strain-aligned part is represented through a modified relaxation rate, while the residual anisotropic part is applied through Guo forcing~\cite{guo2002discrete}. This split is designed to preserve the parts of the stress that are naturally compatible with viscosity-based LBM coupling while retaining non-dissipative contributions that would be lost in a purely eddy-viscosity formulation.

The contributions of this work are as follows. First, we study whether compact macroscopic derivative features provide a sufficient basis for predicting SGS stresses in an LBM setting. Second, we introduce a hybrid stress-splitting strategy that couples the dissipative part of the learned stress through relaxation and the residual part through forcing, enabling stable long-time LES in the cases examined here. Third, we assess the resulting model both a priori and a posteriori, by comparing LES with filtered DNS across stress and energy-transfer p.d.f.s, energy balance, and backscatter statistics. Finally, we show that the trained network can be exported to ONNX Runtime and integrated into a production Fortran/OpenACC LBM code with modest overhead in the tested deployment.

The remainder of the paper is organized as follows. Section~\ref{sec:related} reviews classical SGS modeling approaches and recent ML-based closures, with emphasis on LBM implementations. Section~\ref{sec:methods} describes the governing equations, solver setup, feature design, and training methodology. Section~\ref{sec:results} presents both a priori and a posteriori validation, including ablation studies. Section~\ref{app:implementation} summarizes deployment and performance. Section~\ref{sec:conclusion} summarizes the findings.

\section{Related work}\label{sec:related}

\subsection{Classical SGS models}
A wide range of subgrid–scale (SGS) closures has been proposed for LES. Functional (eddy–viscosity) models approximate the deviatoric SGS stress as
$\tau'_{ij} = -2\nu_t S_{ij}$, with the eddy viscosity $\nu_t$ determined from resolved fields. Classical eddy-viscosity models such as the Smagorinsky model~\cite{smagorinsky1963general} and its dynamic variants~\cite{germano1991dynamic,lilly1992proposed} remain attractive because of their robustness and low cost, but when constrained to nonnegative eddy viscosity they suppress backscatter and have limited ability to represent full stress anisotropy. 

Structural models attempt to reconstruct tensorial features of the SGS stress from the resolved field. Representative examples include the scale–similarity (Bardina) model~\cite{bardina1980improved} and the gradient/Clark model~\cite{clark1979evaluation}. These approaches can capture some non-dissipative behavior but often require additional stabilization (e.g.\ mixed formulations combining similarity/gradient terms with eddy viscosity~\cite{meneveau2000scale}) to ensure numerical robustness and acceptable dissipation levels. Despite extensive refinements, classical closures still face challenges in universality across flow types, Reynolds numbers, filter widths, and grid resolutions.

\subsection{ML–based SGS models}
Data-driven turbulence modeling has developed rapidly across both Reynolds-averaged and LES settings. Representative tensor-valued learning studies showed that neural networks can be trained to predict anisotropic turbulence stresses in channel flow~\cite{fang2020neural}, illustrating the feasibility of learning structured stress tensors rather than only scalar closure coefficients. In LES, a substantial body of work in continuum Navier--Stokes solvers has focused on learning SGS stresses, SGS forces, or related closure terms from resolved flow features~\cite{beck2019deep,bose2023accurate,choi2025perspective,park2021toward,xie2020modeling,yang2025neural}, with a posteriori stability and generalization emerging as central concerns. These models can represent anisotropy and backscatter more flexibly than purely functional eddy-viscosity closures, but strong a priori agreement does not by itself guarantee stable closed-loop LES~\cite{choi2025perspective,park2021toward}.

The broader ML turbulence literature also includes reinforcement-learning approaches to closure discovery. For homogeneous isotropic turbulence (HIT), multi-agent reinforcement learning has been used to discover turbulence closures~\cite{novati2021automating}; related work has extended this perspective to wall-model discovery in wall-bounded turbulence~\cite{bae2022scientific}. We cite these studies because they are important landmarks in data-driven turbulence modeling, even though their learning objectives, solver couplings, and deployment strategies differ from the supervised stress-regression approach adopted here.

Most explicit-SGS-stress ML studies have been developed in continuum Navier--Stokes LES settings rather than in LBM. Across that literature, common inputs include local velocity gradients and derived tensorial quantities such as strain rate, rotation/vorticity, and, in some cases, resolved-stress or Leonard/test-filtered terms; some studies also employ invariant-based or nonlocal feature representations~\cite{choi2025perspective,park2021toward,yang2025neural}.

A closely related example is Yang et al.~\cite{yang2025neural}, who predict the full SGS stress tensor in a Navier--Stokes LES framework using strain-rate and modified Leonard-stress inputs with data-only training. We cite that work as an important continuum-solver reference point; the distinction of the present work lies not in predicting a full stress tensor per se, but in doing so within LBM using derivative-only local features, physics-constrained training, and a dissipative/residual stress split for coupling.

\subsection{LBM–specific LES and learned closures}
A kinetic-theory perspective on turbulence modeling in LBM was introduced by Chen \textit{et al.} in the extended Boltzmann kinetic equation for turbulent flows~\cite{chen2003extended} and further developed in their expanded analogy between Boltzmann kinetic theory and turbulence~\cite{chen2004expanded}. This kinetic-theory-based formulation is a natural target for comparison. Within the LBM, stress and pressure are locally available in space and time from the distribution functions, which makes LBM a natural setting for LES closures that depend on local strain-rate and related quantities. LES has most commonly been realized by modifying relaxation times to represent an eddy viscosity associated with a chosen SGS model~\cite{premnath2009dynamic,zhang2018discrete,zhang2024subgrid}. This viscosity-based strategy fits naturally within LBM and has therefore remained the dominant LES pathway. More recent ML activity in LBM has explored data-driven tuning of relaxation parameters and learned collision operators~\cite{bedrunka2025machine,fischer2025optimal}. Other recent directions include physics-informed near-wall modeling~\cite{xue2024physics} and learned kinetic operators~\cite{xue2025fastforward,corbetta2023toward,ortali2025enhancing}.

Existing studies already show that machine learning is active in LBM, including turbulence-related closures, near-wall modeling, and learned kinetic operators~\cite{xue2024physics,bedrunka2025machine,fischer2025optimal,corbetta2023toward,ortali2025enhancing}. Relative to the LBM studies cited above, the present work focuses on a different question: whether the full SGS stress tensor can be predicted from local derivative features and coupled to LBM through a hybrid dissipative/residual split. In the present formulation, the strain-aligned dissipative part is represented through an effective viscosity and the corresponding stress-mode relaxation-rate modification, while the remaining anisotropic residual is introduced through a forcing term (Guo forcing in our implementation)~\cite{guo2002discrete}. This construction is intended to retain SGS effects that are not naturally represented by a scalar eddy-viscosity closure alone.

\medskip
\noindent\textit{Positioning.} Classical SGS models emphasize robustness, but often at the cost of structural fidelity. The kinetic-theory formulation of Chen \textit{et al.}~\cite{chen2003extended} provides a principled baseline; a central question is whether ML can learn a closure that is more realistic, general, or accurate. ML closures in continuum solvers have shown that explicit stress prediction can recover richer SGS behavior, but their coupling strategies do not transfer directly to LBM. Existing LBM studies, meanwhile, have mostly remained within viscosity-based or kinetic-modification paradigms. The present study is positioned at the intersection of these lines: it examines whether a compact explicit-stress ML closure can be coupled stably and efficiently to LBM through a dissipative/residual split, and whether that strategy improves selected a priori and a posteriori behavior relative to standard viscosity-only baselines in the flows considered here.

\section{Methods}
\label{sec:methods}

This section describes the methodology for developing and deploying ML-based SGS closure models for LBM-based LES, including the LBM formulation, feature design, network architecture, training strategy, and integration with the production solver.

\subsection{Filtered Navier--Stokes equations}
For reference, we introduce the usual LES definition of the SGS stress by filtering the incompressible Navier--Stokes equations~\cite{pope2000turbulent}. Applying a spatial filter $\overline{(\cdot)}$ and defining $\tau_{ij}^{\mathrm{sgs}} = \overline{u_i u_j} - \bar{u}_i \bar{u}_j$, the filtered momentum equation reads
\begin{align}
\frac{\partial \bar{u}_i}{\partial t} + \frac{\partial}{\partial x_j}\left(\bar{u}_i\bar{u}_j\right) &= -\frac{1}{\rho}\frac{\partial \bar{p}}{\partial x_i} + \nu \nabla^2 \bar{u}_i - \frac{\partial \tau_{ij}^{\mathrm{sgs}}}{\partial x_j} + \bar{f}_i.
\end{align}
We emphasize that our simulations are performed with MRT-LBM; the filtered Navier--Stokes form is introduced only to define $\tau_{ij}^{\mathrm{sgs}}$ and its appearance through $-\nabla\cdot\boldsymbol{\tau}^{\mathrm{sgs}}$.

\subsection{MRT-LBM with Guo forcing}
We use a D3Q19 multiple-relaxation-time (MRT) LBM formulation~\cite{d2002multiple}. In compact form, the evolution with a body force $\mathbf{F}$ (Guo forcing~\cite{guo2002discrete}) can be written as
\begin{equation}
\label{eq:mrt_dns}
\begin{split}
f_\alpha(\mathbf{x}+\mathbf{c}_\alpha\Delta t,t+\Delta t) = f_\alpha(\mathbf{x},t)
 &\quad - \left[\mathbf{M}^{-1}\boldsymbol{\Lambda}\left(\mathbf{m}-\mathbf{m}^{eq}\right)\right]_\alpha \\
 &\quad + \Delta t\left[\mathbf{M}^{-1}\left(\mathbf{I}-\frac{\boldsymbol{\Lambda}}{2}\right)\mathbf{m}^{F}\right]_\alpha.
\end{split}
\end{equation}
where $\mathbf{m}=\mathbf{M}\mathbf{f}$ are moments, $\boldsymbol{\Lambda}$ is the diagonal matrix of MRT relaxation rates, and $\mathbf{m}^{F}=\mathbf{M}\boldsymbol{\Phi}(\mathbf{F})$ is the forcing term in moment space. The D3Q19 discrete-velocity set and the corresponding equilibrium distribution are summarized in~\ref{app:d3q19}. Implementation and validation details of the MRT-LBM DNS/LES solver and forcing treatment used in this work can be found in~\cite{khan2025validating}. All DNS, training-data generation, LES baselines (including dynamic Smagorinsky), and ONNX-coupled inference rollouts were performed on an NVIDIA A100 (80~GB) GPU. For deployment we use ONNX Runtime GPU (v1.16.3); the reported runs used CUDA v12.2. All quantities are reported in lattice units (lu).

\subsection{Conventional LBM LES viscosity closure}
In this conventional LBM LES approach, an eddy viscosity $\nu_t$ is introduced through an effective viscosity
\begin{align}
\nu_e &= \nu_0 + \nu_t, \\
\tau_e &= \frac{1}{2} + 3\nu_e, \qquad s_\nu=\frac{1}{\tau_e},
\end{align}
where $\nu_0$ is the molecular viscosity (lu), $\tau_e$ is the effective relaxation time for stress modes, and $s_\nu$ is the corresponding relaxation rate. For further discussion of the viscosity-based LBM LES formulation and its numerical implementation in FHIT, see~\cite{khan2025validating}. For the static Smagorinsky baseline we use $C_s=0.2$; in figure legends, ``Smagorinsky'' denotes this static model and ``Dynamic'' the dynamic procedure below.

\subsubsection{Dynamic Smagorinsky}
To benchmark the ML-LES model against a conventional dynamic eddy-viscosity closure in periodic FHIT, we implement a Germano-type dynamic Smagorinsky procedure in the LBM solver. The deviatoric SGS stress is modeled as
\begin{align}
            \tau'_{ij} &= -2\nu_t\,\bar{S}_{ij},
\qquad
\nu_t = C\,\Delta^2\,|\!\bar{S}\!|,
\qquad
|\!\bar{S}\!|=\sqrt{2\bar{S}_{ij}\bar{S}_{ij}},
\end{align}
where $C\equiv C_s^2$ is obtained dynamically and $\Delta$ is the grid filter width (taken as one lattice spacing in the baseline runs). Consistent with MRT-LBM practice, the strain-rate tensor is evaluated locally from non-equilibrium moments rather than finite differences. Denoting $\mathbf{m}^{neq}=\mathbf{m}-\mathbf{m}^{eq}$ and the current stress-mode relaxation rate by $s_\nu$, the baseline uses an MRT moment-based relation $\bar{S}_{ij}=\mathcal{G}_{ij}(\mathbf{m}^{neq};s_\nu)$.

We define a test filter $\widetilde{(\cdot)}$ with ratio $r\equiv\widetilde{\Delta}/\Delta$ (taken as $r=2$ in our FHIT simulations). The Leonard stress is
\begin{align}
L_{ij} = \widetilde{\bar{u}_i\bar{u}_j} - \widetilde{\bar{u}}_i\,\widetilde{\bar{u}}_j,
\end{align}
and the model tensor is formed as
\begin{align}
M_{ij} = \widetilde{\Delta}^{2}\,|\!\widetilde{\bar{S}}\!|\,\widetilde{\bar{S}}_{ij} - \widetilde{\Delta^{2}\,|\!\bar{S}\!|\,\bar{S}_{ij}}.
\end{align}
The dynamic coefficient is computed using deviatoric contractions,
\begin{align}
L'_{ij} &= L_{ij} - \tfrac{1}{3}L_{kk}\,\delta_{ij},
\qquad
M'_{ij} = M_{ij} - \tfrac{1}{3}M_{kk}\,\delta_{ij},
\qquad
C = -\frac{1}{2}\,\frac{\left\langle L'_{ij}M'_{ij}\right\rangle}{\left\langle M'_{ij}M'_{ij}\right\rangle},
\end{align}
where $\langle\cdot\rangle$ denotes a domain average appropriate for homogeneous turbulence. The resulting eddy viscosity is combined with the molecular value as $\nu_e=\nu_0+\nu_t$ and mapped to the local stress relaxation rate through $\tau_e=\tfrac12+3\nu_e$ and $s_\nu=1/\tau_e$. In the present baseline implementation, $\nu_t$ is clipped to be non-negative before updating $s_\nu$~\cite{premnath2009dynamic}.

\subsection{ML-coupled LBM: effective viscosity and residual forcing}
The ML model predicts the six independent components of the SGS stress tensor; a deviatoric projection (trace removal) is applied post-inference to obtain $\tau^{\mathrm{ML}}_{ij}$. We define the SGS energy transfer channel as
\begin{align}
\Pi = -\tau^{\mathrm{ML}}_{ij} S_{ij},
\end{align}
where $S_{ij}$ is the resolved strain-rate tensor and we use $|\!S\!|=\sqrt{2S_{ij}S_{ij}}$.
We then extract an effective viscosity contribution by dissipation matching \cite{bardina1980improved}:
\begin{align}
\nu_t^{\mathrm{eff}} = \frac{\Pi}{|\!S\!|^2},
\qquad
\nu_e = \nu_0 + \nu_t^{\mathrm{eff}}.
\end{align}
In contrast to common eddy-viscosity baselines, we do not enforce non-negativity of $\Pi$ or $\nu_t^{\mathrm{eff}}$; negative values (backscatter) are therefore retained rather than suppressed by construction.
We then compute the stress relaxation rate from the effective viscosity as $s_\nu=1/(\frac12+3\nu_e)$.
To retain non-dissipative SGS effects, we define the dissipative projection $\tau^{\mathrm{diss}}_{ij}=-2\nu_t^{\mathrm{eff}}S_{ij}$ and the residual stress
\begin{align}
\tau^{\mathrm{res}}_{ij} = \tau^{\mathrm{ML}}_{ij} - \tau^{\mathrm{diss}}_{ij}.
\end{align}
The corresponding residual force density is applied via Guo forcing~\cite{guo2002discrete} for the results reported herein; the exact-difference method~\cite{kupershtokh2009equations} yields similar behavior:
\begin{align}
\mathbf{F}^{\mathrm{res}} = -\nabla\cdot\boldsymbol{\tau}^{\mathrm{res}},
\qquad
\mathbf{F}_{\mathrm{total}} = \mathbf{F}_{\mathrm{ext}} + \mathbf{F}^{\mathrm{res}}.
\end{align}
In practice we evaluate $\nabla\cdot\boldsymbol{\tau}^{\mathrm{res}}$ using centered finite differences with periodic boundary conditions, consistent with the LBM discretization.

\medskip
\noindent\textit{Complete ML--LES LBM update.}
Starting from the DNS MRT-LBM evolution with forcing (Eq.~\ref{eq:mrt_dns}), the ML-coupled LES update is obtained by (i) replacing the stress-mode relaxation rate inside $\boldsymbol{\Lambda}$ by the ML-based value $s_\nu(\nu_e)$, and (ii) using the total forcing $\mathbf{F}_{\mathrm{total}}=\mathbf{F}_{\mathrm{ext}}+\mathbf{F}^{\mathrm{res}}$:
\begin{equation}
\label{eq:mrt_ml_les}
\begin{split}
f_\alpha(\mathbf{x}+\mathbf{c}_\alpha\Delta t,t+\Delta t) = f_\alpha(\mathbf{x},t)
 &\quad - \left[\mathbf{M}^{-1}\boldsymbol{\Lambda}^{\mathrm{ML}}\left(\mathbf{m}-\mathbf{m}^{eq}\right)\right]_\alpha \\
 &\quad + \Delta t\left[\mathbf{M}^{-1}\left(\mathbf{I}-\frac{\boldsymbol{\Lambda}^{\mathrm{ML}}}{2}\right)\mathbf{m}^{F_{\mathrm{total}}}\right]_\alpha.
\end{split}
\end{equation}
where $\boldsymbol{\Lambda}^{\mathrm{ML}}$ equals the DNS relaxation matrix except for the stress relaxation rate $s_\nu$, which is computed from $\nu_e=\nu_0+\nu_t^{\mathrm{eff}}$ with $\nu_t^{\mathrm{eff}}=\Pi/|\!S\!|^2$ and $\Pi=-\tau^{\mathrm{ML}}_{ij}S_{ij}$.

Deployment details for the ONNX-coupled inference pathway are summarized in~\ref{app:implementation}; a step-by-step algorithm for reproducibility is given in Algorithm~\ref{alg:ml_les_inference}.

By construction, the residual stress $\boldsymbol{\tau}^{\mathrm{res}}$ is orthogonal to the strain-rate tensor and therefore non-dissipative. Characterization of the stress split and numerical verification of this property during a posteriori rollouts are presented in Section~\ref{sec:stress_split_results} (Figure~\ref{fig:ml_les_diagnostics}c,f).

\subsection{Training data, features, and preprocessing}
\label{sec:datasets}

Table~\ref{tab:training_cases} summarizes the main training cases. The cases are pooled into a single dataset and split by files into train (80\%), validation (10\%), and test (10\%); all DNS fields are generated with the MRT-LBM solver described in Section~\ref{sec:methods}.

\begin{table}[t]
\centering
\caption{Summary of main training cases used for the reported results. Here $\Delta$ denotes the filter width (in grid points of the DNS), and $\eta^{+}=\eta/\Delta x$ is the dimensionless Kolmogorov length scale.}
\label{tab:training_cases}
\begin{tabular}{lclcc}
\toprule
DNS & $\eta^{+}$ & $\Delta$ & FD grid & Target $\nu$ \\
\midrule
$256^3$ & 0.6 & 4 & $64^3$ & 0.001 \\
$256^3$ & 1.3 & 8 & $32^3$ & 0.004 \\
$192^3$ & 0.8, 1.5 & 6 & $32^3$ & 0.002, 0.005 \\
\bottomrule
\end{tabular}
\end{table}

For the main ML results, the model employs 9 macroscopic derivative features: the six independent components of the symmetric strain-rate tensor
\begin{align}
S_{ij} = \frac{1}{2}\left(\frac{\partial u_i}{\partial x_j} + \frac{\partial u_j}{\partial x_i}\right),
\end{align}
namely $S_{xx}$, $S_{yy}$, $S_{zz}$, $S_{xy}$, $S_{yz}$, $S_{xz}$, together with the three components of the vorticity vector $\boldsymbol{\omega}=\nabla\times\mathbf{u}$, namely $\omega_x$, $\omega_y$, $\omega_z$. For the ML closure, strain-rate components are evaluated locally using centered finite differences on the velocity field, both during data preparation and at inference. To ensure consistent input scaling across datasets generated at different filter widths, we apply a per-simulation $\Delta$-normalization as part of the preprocessing described below.

Derivative-based features such as strain-rate and vorticity scale inversely with the filter width, i.e., $S_{ij},\,\omega_i \sim U/\Delta$. As a consequence, their numerical magnitudes differ across datasets filtered at different resolutions, which would otherwise introduce a spurious dependency on $\Delta$ into the learning process. To make the derivative inputs comparable across filter widths, we apply a per-file (per-simulation) $\Delta$-normalization to all derivative components:
\begin{equation}
S_{ij}^{*} = \Delta^{(s)}\, S_{ij}, \qquad \omega_{i}^{*} = \Delta^{(s)}\, \omega_{i},
\end{equation}
where $\Delta^{(s)}$ is the filter width (equivalently, the downsampling factor) used to generate dataset $s$. In our pipeline, this multiplicative scaling is applied to the derivative features prior to z-score standardization during training, and the same convention is used at inference time.

During inference in LBM--LES, we run on the native lattice without downsampling, so $\Delta^{(s)}=1$ and $S_{ij}^{*}=S_{ij}$ and $\omega_i^{*}=\omega_i$.

\subsubsection{Network Architecture}
\begin{figure*}[t]
    \centering
    \includegraphics[width=0.85\textwidth]{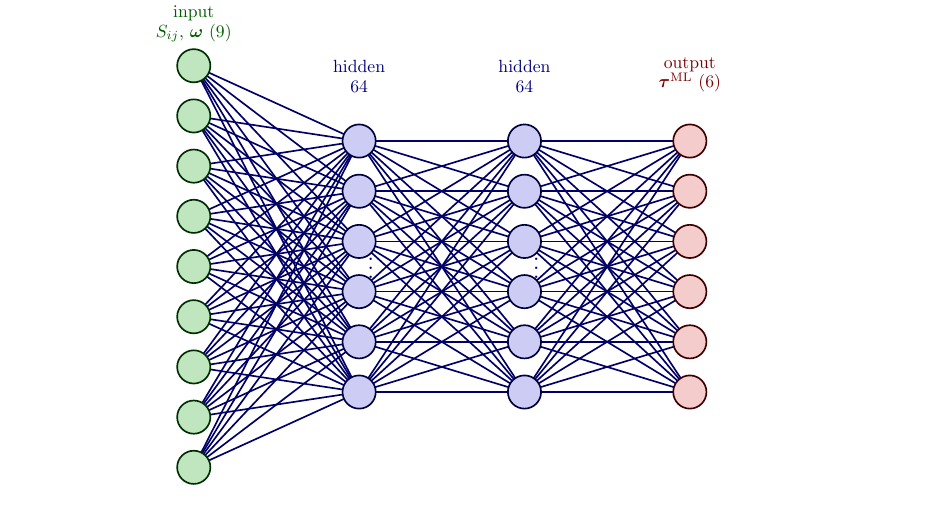}
    \caption{Neural-network architecture for SGS stress prediction (9$\rightarrow$64$\rightarrow$64$\rightarrow$6). A deviatoric projection is applied post-inference.}
    \label{fig:nn_architecture}
\end{figure*}

The core network follows a hierarchical structure: an input linear layer mapping nine features to 64 hidden units, two hidden layers with 64 $\rightarrow$ 64 units, and an output linear layer mapping 64 $\rightarrow$ 6 units for the SGS stress components. We use GELU (Gaussian Error Linear Unit)~\cite{hendrycks2016gaussian} activations for all results reported here.
\begin{align}
    \texttt{GELU}(x) = x \cdot \Phi(x) = x \cdot \frac{1}{2}\left[1 + \text{erf}\left(\frac{x}{\sqrt{2}}\right)\right],
\end{align}
where $\Phi(x)$ is the cumulative distribution function of the standard normal distribution.
The complete model contains approximately 5,190 trainable parameters: Layer~1 has $9\times 64$ weights plus 64 bias parameters, Layer~2 has $64\times 64$ weights plus 64 bias parameters, and the output head has $64\times 6$ weights plus 6 bias parameters, for a total of $640 + 4{,}160 + 390 = 5{,}190$ trainable parameters.

\subsection{Losses: data and physics}
Training combines a data loss on the SGS stress with three physics-constrained losses used for the main results:
\begin{align}
\label{eq:loss_total}
\mathcal{L} = \mathcal{L}_{\text{MSE}} + \lambda_{\pi}\mathcal{L}_{\pi} + \lambda_{\text{eq}}\mathcal{L}_{\text{eq}} + \lambda_{\text{divF}}\mathcal{L}_{\text{divF}}.
\end{align}
\noindent Here $\mathcal{L}_{\text{MSE}}$ is the mean-squared error on the six SGS stress components. The dissipation (energy transfer) channel is defined as
\begin{align}
\Pi_{\text{pred}}^{(i)} = \mathbf{c}^{(i)T}\boldsymbol{\tau}_{\text{pred}}^{(i)}, \quad \mathbf{c}^{(i)} = [-S_{xx}^{(i)}, -S_{yy}^{(i)}, -S_{zz}^{(i)}, -2S_{xy}^{(i)}, -2S_{xz}^{(i)}, -2S_{yz}^{(i)}]^T.
\end{align}
\noindent The dissipation-matching loss is
\begin{align}
\label{eq:loss_pi}
\mathcal{L}_{\pi}=\mathrm{MSE}(\Pi_{\text{pred}},\Pi_{\text{true}}),
\end{align}
which preserves natural backscatter (no enforced non-negativity hinge). The equivariance loss $\mathcal{L}_{\text{eq}}$ enforces rotational consistency under the discrete cube-rotation group, i.e.\ $\tau(Rx)\approx R\tau(x)R^T$. The divergence-free SGS-force loss $\mathcal{L}_{\text{divF}}$ penalizes $\nabla\cdot\mathbf{F}_{\text{sgs}}$ where $\mathbf{F}_{\text{sgs}}=-\nabla\cdot\boldsymbol{\tau}'$, encouraging numerical compatibility with the forcing formulation. Because $\mathcal{L}_{\text{divF}}$ requires spatial derivatives of the predicted stress, it is evaluated on full-grid snapshots rather than within the shuffled batch loop; we compute it once per epoch on a subset of training files and apply a separate gradient step, while the data loss, $\Pi$-matching, and equivariance loss are applied per batch.
\medskip
\noindent For the main-results model we use $\lambda_{\pi}=0.4$, $\lambda_{\text{eq}}=0.1$, and $\lambda_{\text{divF}}=0.05$. No clipping or value-limiting is applied to SGS-stress targets or network outputs during training, nor to predicted stresses, $\Pi$, $\nu_t^{\mathrm{eff}}$, or the resulting SGS forcing during a posteriori rollouts.

\section{Results}
\label{sec:results}

This section presents a priori validation, ablation studies, and a posteriori LES results across the parameter regimes and flow conditions considered. Unless otherwise stated, time is reported in normalized form as $t^{*}=t/t_{0}$ with $t_{0}=L_{0}/U_{0}$ ($L_{0}$ domain length, $U_{0}=0.043$ reference velocity in lu; $t^{*}=\mathrm{iter}\,U_{0}/N_{x}$ with $\Delta x=\Delta t=1$). Statistics are computed from snapshots with $t^{*}\geq 20$ where the flow is statistically stationary (Fig.~\ref{fig:energy_balance}). All ML-based and Smagorinsky baseline results are for $128^3$ at $\nu=0.001$ (lu).

\subsection{A priori validation}

We assess model performance across the training parameter space using stress-correlation metrics and physical consistency checks.

As a primary a priori metric for stress prediction accuracy, we report the Pearson correlation coefficient between predicted and FD SGS stress components. For each independent component $k\in\{xx,yy,zz,xy,xz,yz\}$, we compute
\begin{align}
\mathrm{Corr}(\tau_{k,\text{pred}}, \tau_{k,\text{true}})
&= \frac{\mathrm{Cov}(\tau_{k,\text{pred}}, \tau_{k,\text{true}})}{\sigma_{\tau_{k,\text{pred}}}\,\sigma_{\tau_{k,\text{true}}}} .
\end{align}
We report correlations per component (and, when needed, aggregate across components).

\begin{figure}[t]
    \centering
    \includegraphics[width=0.95\linewidth]{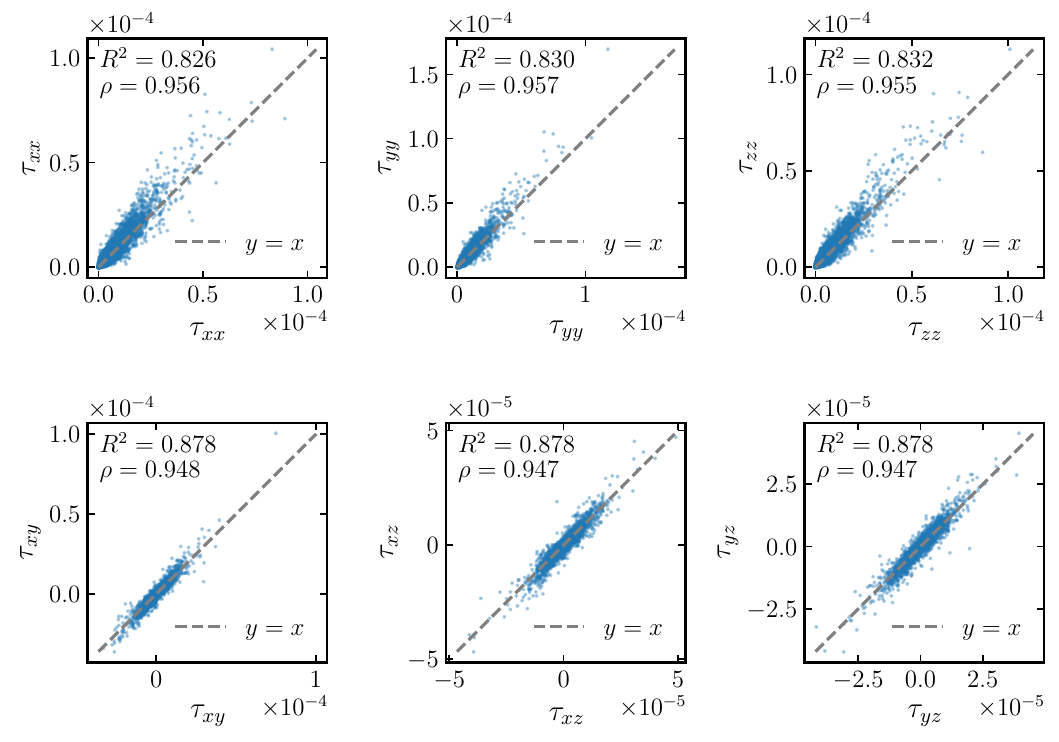}
    \caption{A priori validation scatter plots for SGS stress components. Each panel compares the predicted SGS stress component $\tau_{ij}^{\mathrm{pred}}$ against the FD reference $\tau_{ij}^{\mathrm{true}}$ over a validation dataset: FD grid $128^3$ ($256^3$ DNS, $\Delta=2$), $\nu=0.001$. The dashed line indicates $y=x$. Reported values are the coefficient of determination $R^2$ and Pearson correlation $\rho$ computed per component.}
    \label{fig:stress_scatter_priori}
\end{figure}

Figure~\ref{fig:stress_scatter_priori} shows representative a priori scatter plots comparing predicted versus FD SGS stress components; annotations report $R^2$ and Pearson correlation $\rho$ for each component.

\begin{figure}[t]
    \centering
    \includegraphics[width=0.75\linewidth]{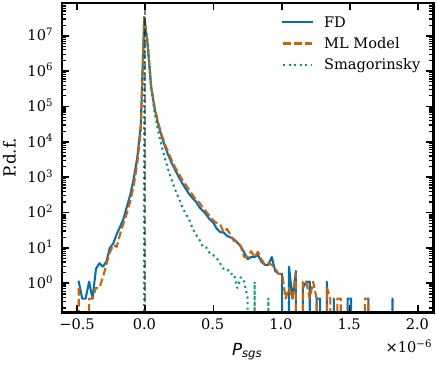}
    \caption{A priori p.d.f. of the SGS energy transfer $\Pi$ comparing FD reference, ML prediction, and static Smagorinsky baseline (a priori). The ML model closely matches FD across the peak and tails, including backscatter (negative $\Pi$) and intermittent forward transfer (positive $\Pi$). For Smagorinsky, the non-negative eddy viscosity yields zero p.d.f. for $\Pi<0$ and underpredicts the positive-$\Pi$ tail.}
    \label{fig:psgs_pdf_priori}
\end{figure}

To assess whether the model reproduces the distribution of SGS energy transfer and intermittent events~\cite{borue1998local}, we also compare the probability density function (p.d.f.) of the SGS energy transfer proxy defined from the stress and strain-rate as
\begin{align}
\Pi = -\tau_{ij} S_{ij}.
\end{align}
Figure~\ref{fig:psgs_pdf_priori} compares the p.d.f. of $\Pi$ from the ML prediction and the FD reference. For completeness, we also include the a priori p.d.f. of the static Smagorinsky baseline (evaluated on the same FD fields). The negative tail corresponds to backscatter events (net transfer from subgrid to resolved scales), while positive values correspond to forward transfer. The ML model closely matches the FD reference across the entire distribution, including the intermittent backscatter and forward-transfer tails. In contrast, the Smagorinsky model, by construction, enforces non-negative eddy viscosity and thus yields zero p.d.f. for $\Pi<0$ and significantly underpredicts the probability of larger positive-$\Pi$ events.

To quantify how stress-prediction accuracy varies across the training parameter space, we evaluate the model on each $(\Delta,\nu)$ bucket of the full dataset from the training configurations (Table~\ref{tab:training_cases}). Table~\ref{tab:bucket_metrics} reports the coefficient of determination $R^2$ and macro-averaged Pearson correlation $\rho$ per bucket. The $\Delta$-normalization (Section~\ref{sec:datasets}) ensures comparable feature scaling across buckets. Performance is strongest for $(\Delta=6,\nu=0.005)$ and $(\Delta=8,\nu=0.004)$, with $R^2\approx 0.80$--$0.84$ and $\rho\approx 0.90$--$0.95$; the $(\Delta=4,\nu=0.001)$ bucket, which contributes the largest sample count, yields $R^2\approx 0.70$ and $\rho\approx 0.84$. This per-bucket breakdown confirms that the model achieves consistent accuracy across the full range of filter widths and viscosities used in training.

\begin{table}[t]
\centering
\caption{$R^2$ and Pearson correlation $\rho$ per $(\Delta,\nu)$ bucket on the full dataset from the training configurations (Table~\ref{tab:training_cases}). $N$ is the number of samples per bucket.}
\label{tab:bucket_metrics}
\begin{tabular}{ccclcc}
\toprule
$\Delta$ & $\nu$ & FD grid & $N$ & $R^2$ & $\rho$ \\
\midrule
4 & 0.001 & $64^3$ & 13.4M & 0.696 & 0.838 \\
6 & 0.002 & $32^3$ & 1.7M & 0.710 & 0.844 \\
6 & 0.005 & $32^3$ & 1.7M & 0.837 & 0.952 \\
8 & 0.004 & $32^3$ & 1.7M & 0.804 & 0.904 \\
\bottomrule
\end{tabular}
\end{table}

The results reveal a clear trend: accuracy improves with higher viscosity and coarser filter width. At fixed $\Delta=6$, increasing $\nu$ from 0.002 to 0.005 raises $R^2$ from 0.71 to 0.84; similarly, $(\Delta=8,\nu=0.004)$ outperforms $(\Delta=4,\nu=0.001)$. The most challenging regime is the finest filter with lowest viscosity $(\Delta=4,\nu=0.001)$, which also contributes the largest sample count. This pattern is consistent with the expectation that more dissipative flows (higher $\nu$) and coarser filters (larger $\Delta$) yield more regular SGS structure and thus more predictable stress fields.

\subsection{A posteriori}

\subsubsection{Characterization and verification of the stress split}
\label{sec:stress_split_results}
\begin{figure}[t]
    \centering
    \includegraphics[width=0.95\linewidth]{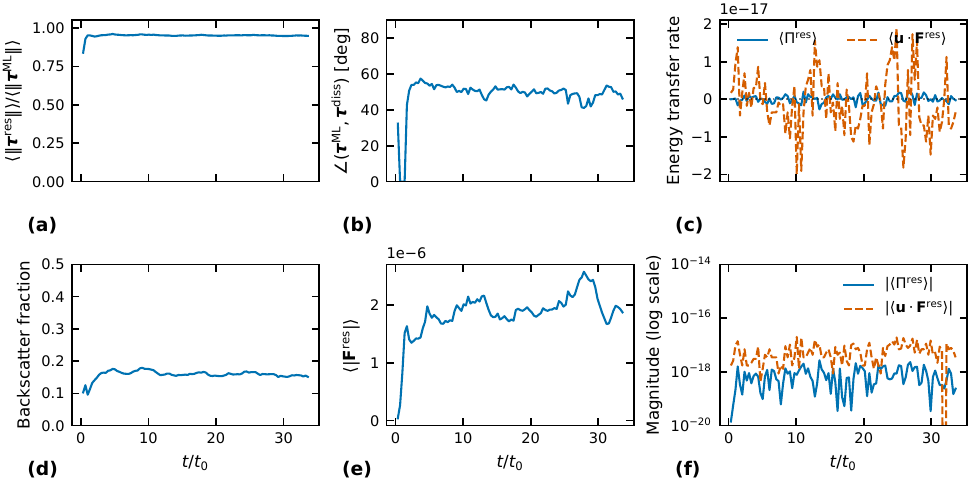}
    \caption{Characterization and verification of the hybrid SGS stress-splitting closure versus normalized time $t/t_0$.
    (a) Residual fraction $\langle\|\boldsymbol{\tau}^{\mathrm{res}}\|\rangle / \langle\|\boldsymbol{\tau}^{\mathrm{ML}}\|\rangle$, quantifying the portion of the learned SGS stress not representable by an eddy-viscosity-aligned contribution.
    (b) Global stress-alignment proxy between the learned stress and its dissipative component, computed from domain-averaged inner products and magnitudes.
    (c) Energy consistency: residual dissipation $\Pi^{\mathrm{res}} = -\tau^{\mathrm{res}}_{ij} S_{ij}$ and resolved-scale work of the residual forcing $\langle \mathbf{u}\cdot\mathbf{F}^{\mathrm{res}}\rangle$, with $\mathbf{F}^{\mathrm{res}}=-\nabla\cdot\boldsymbol{\tau}^{\mathrm{res}}$.
    (d) Backscatter fraction, defined as the fraction of grid cells where $\Pi=-\tau^{\mathrm{ML}}_{ij}S_{ij}<0$.
    (e) Mean magnitude of the residual forcing $\langle|\mathbf{F}^{\mathrm{res}}|\rangle$, indicating the activity level of the explicit anisotropic SGS contribution.
    (f) Discrete orthogonality shown on a logarithmic scale; both $|\langle\Pi^{\mathrm{res}}\rangle|$ and $|\langle\mathbf{u}\cdot\mathbf{F}^{\mathrm{res}}\rangle|$ remain at numerical noise levels.
    Panels (c) and (f) demonstrate that the residual stress is non-dissipative and that its explicit injection via forcing is energy-consistent at the discrete level. Data from a representative FHIT rollout ($128^3$, $\nu=0.001$).}
    \label{fig:ml_les_diagnostics}
\end{figure}

Figure~\ref{fig:ml_les_diagnostics} summarizes both the structure and the numerical consistency of the proposed SGS closure during the coupled LES rollout. Panels (a) and (b) show that a substantial fraction of the learned SGS stress is not aligned with the strain-rate tensor and therefore cannot be represented by a purely eddy-viscosity-based model. Panel (e) confirms that the corresponding residual forcing is dynamically active. Panels (c) and (f) provide a discrete verification of the formulation: the residual stress is orthogonal to the strain rate and its explicit injection via forcing does not introduce net resolved-scale energy transfer, with both measures remaining at machine precision. Together, these results demonstrate a clean separation between dissipative and anisotropic SGS contributions.

\subsubsection{Robustness across collision operators}
\begin{figure}[t]
    \centering
    \includegraphics[width=0.75\linewidth]{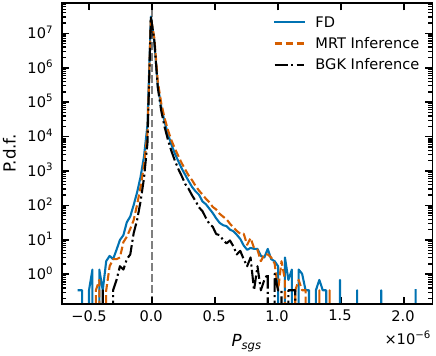}
    \caption{A posteriori p.d.f. of $\Pi$ comparing FD reference with ML inference deployed in LBM with MRT and BGK collision operators. The BGK configuration remains statistically consistent with FD, indicating transferability beyond the collision model used to generate the training data.}
    \label{fig:psgs_mrt_bgk}
\end{figure}

To probe robustness with respect to the collision operator, we compare a posteriori statistics of $\Pi$ obtained when the same trained network is coupled to LBM with either MRT or Bhatnagar--Gross--Krook (BGK) collision operators. Figure~\ref{fig:psgs_mrt_bgk} shows that both configurations closely follow the FD reference, including the negative-$\Pi$ tail (backscatter), even though the training dataset was generated from MRT-LBM DNS.

\begin{figure}[t]
    \centering
    \begin{minipage}[t]{0.49\linewidth}
        \includegraphics[width=\linewidth]{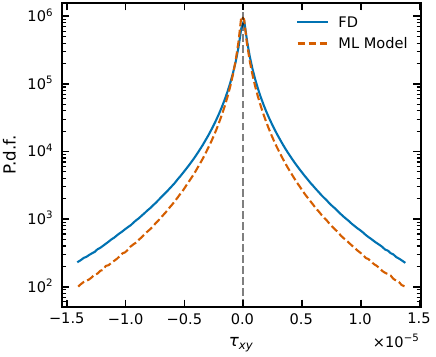}\\[2pt]
        \raggedright \textbf{(a)}
    \end{minipage}\hfill
    \begin{minipage}[t]{0.49\linewidth}
        \includegraphics[width=\linewidth]{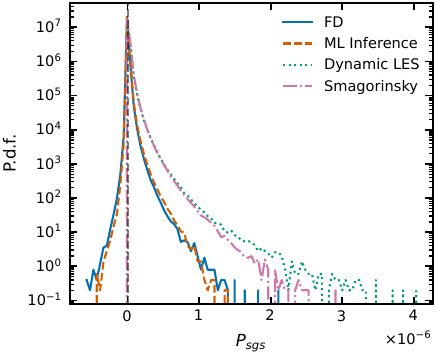}\\[2pt]
        \raggedright \textbf{(b)}
    \end{minipage}
    \caption{A posteriori p.d.f. comparisons: (a) representative SGS stress component distribution, and (b) p.d.f. of $\Pi$ comparing FD, ML inference, and classical SGS baselines.}
    \label{fig:posteriori_pdfs}
\end{figure}

Figure~\ref{fig:posteriori_pdfs} summarizes a posteriori probability density function (p.d.f.) comparisons computed from the coupled LES rollouts. Panel (a) compares the p.d.f. of a representative SGS stress component and shows that the ML-based closure closely follows the FD reference distribution, indicating that the rollout reproduces the typical stress magnitudes and intermittency level of the reference case. Panel (b) compares the p.d.f. of $\Pi$. The ML inference remains well aligned with the FD curve across the core of the distribution and into the tails. In contrast, the eddy-viscosity baselines (dynamic LES and static Smagorinsky) deviate strongly from FD, with distributions biased toward large positive $\Pi$ and an essentially absent negative-$\Pi$ tail due to the strictly non-negative eddy viscosity ($\nu_t \ge 0$) used in the baseline implementations, which suppresses backscatter.

\begin{figure}[t]
    \centering
    \begin{minipage}[t]{\linewidth}
        \centering
        \includegraphics[width=0.95\linewidth]{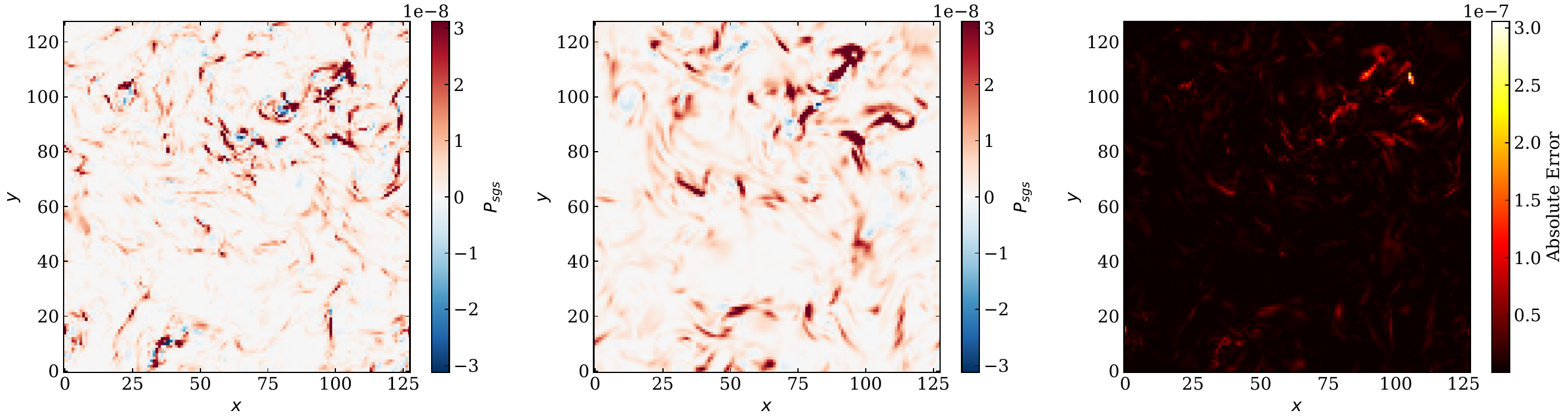}\\[4pt]
        \raggedright \textbf{(a)} ML-LES vs FD
    \end{minipage}\\[8pt]
    \begin{minipage}[t]{\linewidth}
        \centering
        \includegraphics[width=0.95\linewidth]{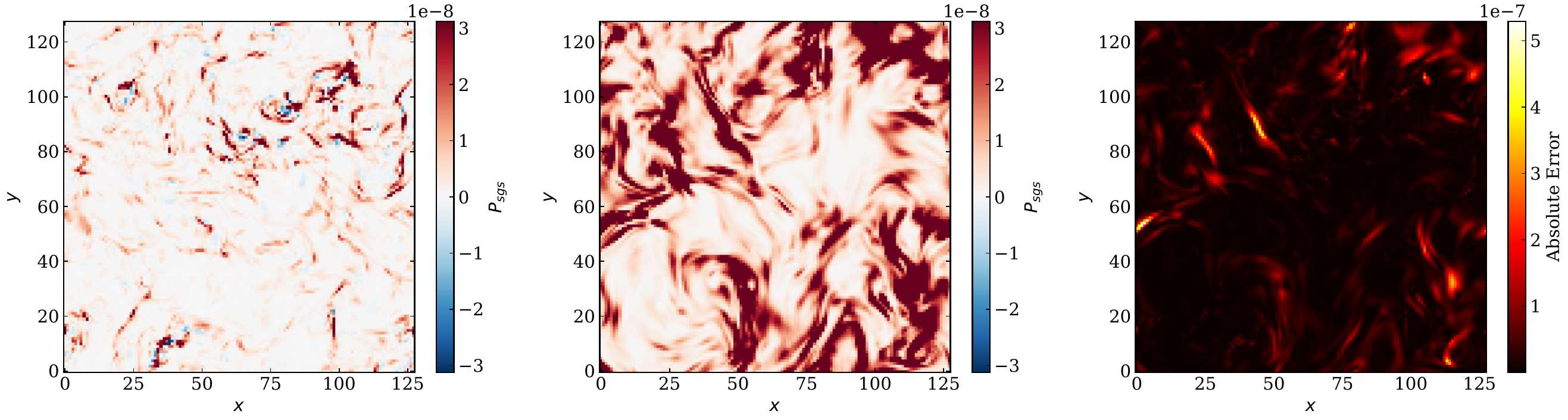}\\[4pt]
        \raggedright \textbf{(b)} Dynamic Smagorinsky vs FD
    \end{minipage}
    \caption{Spatial slice comparison of SGS energy transfer $\Pi$ at $z$-slice 64/128, iteration 89,000 (time step index). In both panels, from left to right: FD (ground truth), model prediction, and absolute error map. (a) ML-LES vs FD. (b) Dynamic Smagorinsky vs FD. The ML closure reproduces the spatial structure of $\Pi$ with smaller error than the dynamic baseline.}
    \label{fig:psgs_spatial}
\end{figure}

Figure~\ref{fig:psgs_spatial} provides a spatial comparison of the SGS energy transfer $\Pi$ at a representative $z$-slice (iteration 89,000). The top row compares ML-LES against the FD reference; the bottom row compares dynamic Smagorinsky against the same FD reference. The ML closure reproduces the spatial structure of $\Pi$ (filamentary patterns, positive and negative regions) with smaller error than the dynamic baseline, whose error map shows larger and more widespread discrepancies.

\begin{figure}[t]
    \centering
    \begin{minipage}[t]{0.49\linewidth}
        \includegraphics[width=\linewidth]{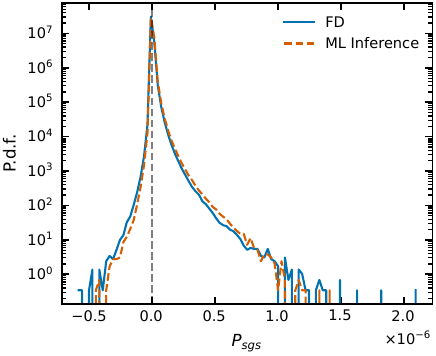}\\[2pt]
        \raggedright \textbf{(a)}
    \end{minipage}\hfill
    \begin{minipage}[t]{0.49\linewidth}
        \includegraphics[width=\linewidth]{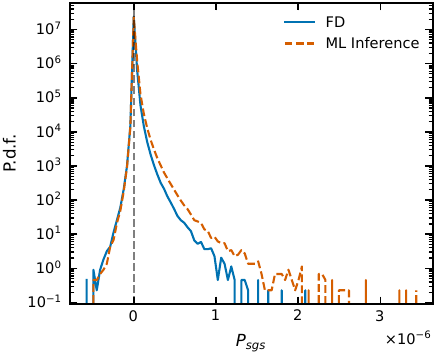}\\[2pt]
        \raggedright \textbf{(b)}
    \end{minipage}
    \caption{A posteriori $\Pi$ p.d.f. from LES rollouts for two ML variants: (a) with physics loss, (b) without physics loss.}
    \label{fig:posteriori_psgs_ablation}
\end{figure}

Figure~\ref{fig:posteriori_psgs_ablation} compares a posteriori $\Pi$ statistics for two training objectives. With the full objective (data loss plus physics losses; Eq.~\ref{eq:loss_total}), the predicted $\Pi$ p.d.f. aligns closely with FD. When all physics losses are disabled and training uses only the stress data loss $\mathcal{L}_{\mathrm{MSE}}$, the model still reproduces the negative-$\Pi$ tail (backscatter), but shows a noticeable deviation for positive $\Pi$, indicating overly strong forward-transfer events.

\begin{figure}[t]
    \centering
    \includegraphics[width=0.75\linewidth]{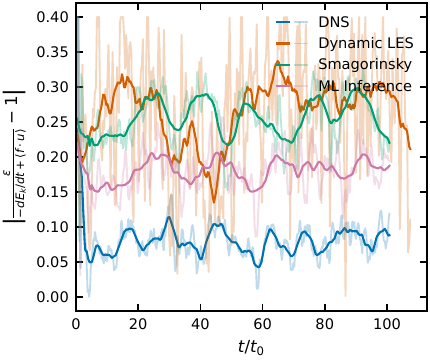}
    \caption{Energy-balance error versus normalized time $t/t_0$: $|\,\varepsilon/(-dK/dt+\langle \mathbf{f}\cdot\mathbf{u}\rangle)-1\,|$ for DNS ($256^3$) and LES rollouts ($128^3$). ML inference stays closest to DNS, while dynamic Smagorinsky shows the largest imbalance.}
    \label{fig:energy_balance}
\end{figure}

To assess whether the closures reproduce correct global energetics in closed-loop LES, we monitor the instantaneous energy-budget mismatch based on the volume-averaged kinetic-energy equation, i.e. the deviation of $\varepsilon$ from the balance $-dK/dt+\langle \mathbf{f}\cdot\mathbf{u}\rangle$. Figure~\ref{fig:energy_balance} shows that the ML-coupled LES remains significantly closer to the DNS reference ($256^3$) than the Smagorinsky baselines on the $128^3$ grid. In particular, the dynamic Smagorinsky case exhibits the largest imbalance, indicating that its viscosity-based dissipation does not track the resolved energy evolution as accurately as the learned explicit-stress closure.

The $128^3$ rollouts above extend beyond the exact training configuration: the training data use FD grids of size $64^3$ and $32^3$ (Table~\ref{tab:training_cases}), whereas inference runs on the native LES grid at $128^3$ without downsampling. To probe stability at even larger resolutions, we run ML inference on a $512^3$ grid. After an initial transient, the energy-balance error stabilizes around 10--12\% (see~\ref{app:energy_balance_512}), indicating that the block-based inference (Algorithm~\ref{alg:ml_les_inference}) remains numerically stable and maintains approximate energy consistency at production-scale resolutions.

As a preliminary a posteriori test of transfer beyond FHIT, we deploy the same trained model in turbulent channel flow; the results are reported in~\ref{app:channel_generalization}.

\section{Conclusion}
\label{sec:conclusion}

We presented a physics-constrained ML-based SGS closure for LBM LES that predicts the SGS stress tensor from local derivative features and couples it through a dissipative/residual split. The validation reported in Sections~\ref{sec:results} and~\ref{app:channel_generalization} suggests that the closure improves upon Smagorinsky baselines in FHIT (a priori and a posteriori) and remains consistent across MRT and BGK collision operators, transfers to channel flow without retraining, and achieves throughput comparable to a dynamic Smagorinsky baseline when deployed via ONNX.

The main contribution is to show that an explicit-stress ML closure can be integrated into LBM in a practical and stable way, with the split mechanism enabling non-dissipative SGS effects that viscosity-only closures suppress. Kang \textit{et al.}~\cite{Kang_Jeon_You_2023} established that some cross-flow transfer from FHIT to channel flow is possible, though only partially in their formulation. Our LBM results show transfer in a formulation that uses $\Delta$-normalized features and a dissipative/residual stress split~(\ref{app:channel_generalization}).

Validation is limited to FHIT (primary) and a preliminary channel-flow transfer test. Future work should clarify the mechanism behind the observed transfer through a rigorous analysis of feature-space overlap and prediction accuracy across flow regimes, extend the assessment to additional flow classes and Reynolds numbers, report skin-friction metrics for channel flow, and explore multi-GPU scaling for larger problem sizes.
\appendix

\section{D3Q19 velocity set and equilibrium distribution}
\label{app:d3q19}
For completeness, we summarize the discrete-velocity stencil and standard second-order equilibrium used by the D3Q19 LBM referenced in Section~\ref{sec:methods}. The D3Q19 velocity set consists of 19 discrete velocity vectors
\begin{align}
\mathbf{c}_\alpha = c\,(c_{\alpha x},c_{\alpha y},c_{\alpha z})^{T},
\end{align}
where $c=\Delta x/\Delta t$ is the lattice speed and $(c_{\alpha x},c_{\alpha y},c_{\alpha z})$ are the dimensionless direction vectors. The velocity vectors are
\begin{align}
\mathbf{c}_\alpha &= c\begin{cases}
(0,0,0) & \alpha=1,\\
(\pm 1,0,0),\ (0,\pm 1,0),\ (0,0,\pm 1) & \alpha=2,\ldots,7,\\
(\pm 1,\pm 1,0),\ (\pm 1,0,\pm 1),\ (0,\pm 1,\pm 1) & \alpha=8,\ldots,19.
\end{cases}
\end{align}
The corresponding lattice weights are $w_0=1/3$ (rest), $w_s=1/18$ (axial), and $w_d=1/36$ (diagonal), and the lattice sound speed satisfies $c_s^2=c^2/3$.

The equilibrium distribution is taken in the standard second-order form
\begin{equation}
f_\alpha^{\mathrm{eq}} = w_\alpha\,\rho\left[1 + \frac{\mathbf{c}_\alpha\!\cdot\!\mathbf{u}}{c_s^2}
+ \frac{(\mathbf{c}_\alpha\!\cdot\!\mathbf{u})^2}{2c_s^4}
- \frac{\mathbf{u}\!\cdot\!\mathbf{u}}{2c_s^2}\right].
\end{equation}
All simulations reported in this work are performed in a cubic domain with periodic boundary conditions in all three directions.

\section{Transfer to channel flow}
\label{app:channel_generalization}
We perform an a posteriori test by deploying the same trained model (Section~\ref{sec:methods}, Table~\ref{tab:training_cases}) in turbulent channel flow at $\mathrm{Re}_\tau \approx 160$. The model was trained exclusively on forced homogeneous isotropic turbulence (FHIT); applying it without retraining to a wall-bounded flow with strong mean shear tests transfer to flow configurations not seen during training---a recognized challenge in ML-based LES~\cite{choi2025perspective}. Unlike prior physics-informed LBM work for wall-bounded flows, which required neural networks trained on near-wall data~\cite{xue2024physics}, our bulk-trained closure is deployed directly with wall boundary conditions. The simulation uses the hybrid ML--LES coupling on a $1536 \times 64 \times 128$ grid (streamwise $\times$ spanwise $\times$ wall-normal), viscosity $\nu=0.001$, and uniform streamwise forcing; statistics are time-averaged after an initial transient.

Kang \textit{et al.}~\cite{Kang_Jeon_You_2023} showed that transfer from FHIT-trained artificial neural network (ANN) closures to channel flow is possible but limited, and argued that this depends importantly on the model inputs, especially the inclusion of the Leonard stress $L_{ij}$. In contrast, our LBM closure transfers without retraining, suggesting that generalization depends not only on training data but also on representation, normalization, and solver coupling. Our ablation without $\Delta$-normalization gave poorer inference results, suggesting that, in our formulation, $\Delta$-normalization contributes to making inputs more comparable across filter widths and flow regimes (Section~\ref{sec:methods}). One interpretation is that the closure depends only on local gradient information---the strain-rate and vorticity at each point---and does not require knowledge of the global flow geometry or wall position. It therefore approximates a local constitutive relation between filtered derivatives and SGS stress, rather than a flow-specific model. Transfer may then occur wherever the distribution of these local features in channel flow overlaps with that encountered in FHIT, despite the different large-scale flow structure. This interpretation remains speculative; rigorous support would require a quantitative analysis of feature-space overlap and its relation to prediction accuracy.

\begin{figure}[t]
    \centering
    \begin{minipage}[t]{0.49\linewidth}
        \includegraphics[width=\linewidth]{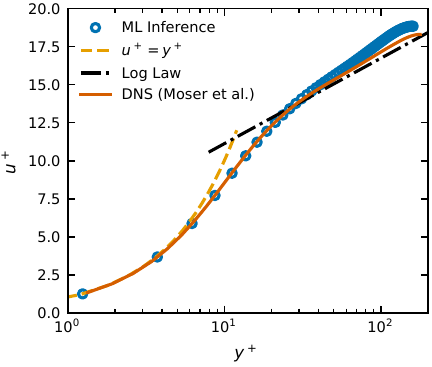}\\[2pt]
        \raggedright \textbf{(a)}
    \end{minipage}\hfill
    \begin{minipage}[t]{0.49\linewidth}
        \includegraphics[width=\linewidth]{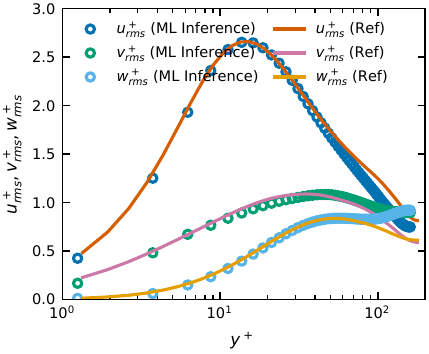}\\[2pt]
        \raggedright \textbf{(b)}
    \end{minipage}\\[8pt]
    \begin{minipage}[t]{0.75\linewidth}
        \centering
        \includegraphics[width=\linewidth]{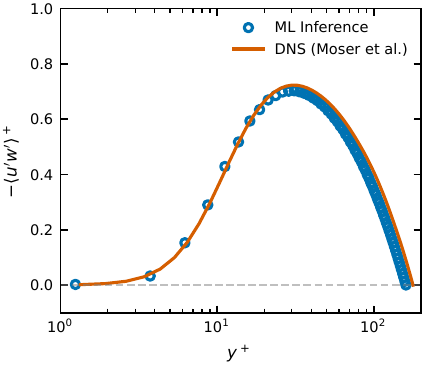}\\[2pt]
        \raggedright \textbf{(c)}
    \end{minipage}
    \caption{A posteriori channel flow at $\mathrm{Re}_\tau \approx 160$ (ML simulation) vs DNS benchmark at $\mathrm{Re}_\tau \approx 180$~\cite{kim1987turbulence}: (a) mean streamwise velocity $u^+$ versus $y^+$; (b) RMS fluctuations $u_{\mathrm{rms}}^+$, $v_{\mathrm{rms}}^+$, $w_{\mathrm{rms}}^+$; (c) Reynolds shear stress $-\langle u'w' \rangle^+$ versus $y^+$ ($w$ wall-normal).}
    \label{fig:channel_generalization}
\end{figure}

Figure~\ref{fig:channel_generalization} compares the resulting mean velocity, RMS fluctuation, and Reynolds shear stress profiles against the DNS benchmark of Kim \textit{et al.}~\cite{kim1987turbulence} at $\mathrm{Re}_\tau \approx 180$; our simulation is at $\mathrm{Re}_\tau \approx 160$. Panel (a) shows the mean streamwise velocity $u^+$ versus $y^+$; the ML-coupled LES shows close agreement with the benchmark and captures both the viscous sublayer and the logarithmic region despite the slight Reynolds-number mismatch. Panel (b) shows the RMS fluctuations $u_{\mathrm{rms}}^+$, $v_{\mathrm{rms}}^+$, and $w_{\mathrm{rms}}^+$, with good agreement across all three components. Panel (c) shows the normalized Reynolds shear stress $-\langle u'w' \rangle^+$ versus $y^+$ ($w$ wall-normal in our convention); the ML closure reproduces the benchmark profile closely across the near-wall, buffer, and logarithmic regions.

\section{Stability at production-scale resolution}
\label{app:energy_balance_512}
\begin{figure}[t]
    \centering
    \includegraphics[width=0.75\linewidth]{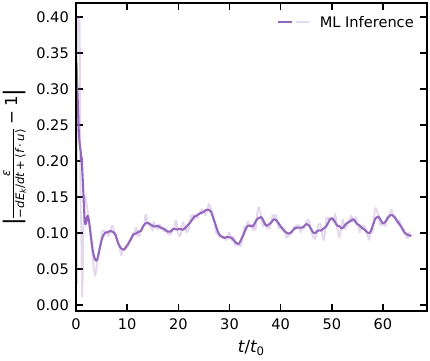}
    \caption{Energy-balance error $|\,\varepsilon/(-dK/dt+\langle \mathbf{f}\cdot\mathbf{u}\rangle)-1\,|$ for ML inference at $512^3$ ($\nu=0.001$, $t/t_0 \approx 60$). Error stabilizes after the transient.}
    \label{fig:energy_balance_512}
\end{figure}

Figure~\ref{fig:energy_balance_512} shows the energy-balance error for ML inference at $512^3$, referenced in Section~\ref{sec:results}. The error stabilizes after the transient, supporting stable deployment at production-scale resolutions.

\section{Implementation and computational performance}
\label{app:implementation}

The ML model is deployed in a production Fortran/OpenACC MRT-LBM solver using ONNX Runtime~\cite{onnxruntime}. We summarize the inference pathway at a high level for reproducibility. At each time step, the solver computes a compact feature vector at every lattice site and evaluates the trained network to obtain the SGS stress used by the ML--LES closure. The deployed model uses nine macroscopic derivative features: the six independent strain-rate components and three vorticity components, ordered as $S_{xx}$, $S_{yy}$, $S_{zz}$, $S_{xy}$, $S_{xz}$, $S_{yz}$, and $\omega_x$, $\omega_y$, $\omega_z$.
The derivatives are evaluated on the periodic lattice using centered finite differences, and the same preprocessing convention as training is applied through a filter-width scaling based on the downsampling factor $\Delta$. In the coupled LES rollouts shown in this paper we run on the native lattice so $\Delta=1$.

Inference is executed through an ONNX Runtime session created once at initialization and reused across time steps. The exported ONNX model embeds z-score normalization of features and denormalization of targets in the graph, so the solver passes raw feature values and receives physical stress values directly. The feature tensor is assembled in Fortran and passed to the inference engine; for GPU execution, the implementation avoids repeated allocations by reusing persistent buffers. For larger problem sizes, the inference call can be executed in batches to cap peak memory while maintaining throughput.

The network outputs the six independent components of the SGS stress tensor. Before coupling to the LBM update, we enforce a deviatoric projection by removing the trace of the predicted stress so that the injected stress is traceless by construction; the original (pre-projection) outputs can optionally be saved for post-processing.

Algorithm~\ref{alg:ml_les_inference} summarizes the per-time-step procedure for reproducibility.

\begin{algorithm}[H]
\caption{ML--LES inference and coupling (per time step)}
\label{alg:ml_les_inference}
\footnotesize
\begin{algorithmic}[1]
\Require Resolved velocity $\mathbf{u}$, density $\rho$; molecular viscosity $\nu_0$; filter width $\Delta$ (unity on native lattice)
\Ensure Updated distribution functions $f_\alpha$ via MRT collision with ML-based closure
\State \textbf{1. Compute features} at all lattice sites:
\State \quad Using velocity $\tilde{\mathbf{u}} = \mathbf{u} + \frac{1}{2}\rho^{-1}\mathbf{F}_{\mathrm{ext}}$ (Guo half-force correction~\cite{guo2002discrete}), evaluate strain-rate $S_{ij} = \frac{1}{2}(\partial_j \tilde{u}_i + \partial_i \tilde{u}_j)$ and vorticity $\boldsymbol{\omega} = \nabla\times\tilde{\mathbf{u}}$ using centered finite differences
\State \quad Apply $\Delta$-normalization: $S^*_{ij} \leftarrow \Delta\, S_{ij}$, $\omega^*_i \leftarrow \Delta\, \omega_i$
\State \quad Assemble feature vector $\mathbf{x} = (S_{xx}, S_{yy}, S_{zz}, S_{xy}, S_{xz}, S_{yz}, \omega_x, \omega_y, \omega_z)$
\State \textbf{2. Run network inference} to obtain raw stress $\boldsymbol{\tau}^{\mathrm{raw}}$ ($N$ = number of lattice sites):
\If{$N > 200{,}000$}
\State run in batches of size $200{,}000$
\ElsIf{$N > 10{,}000$}
\State run in batches of size $50{,}000$
\Else
\State run full-field inference
\EndIf
\State \textbf{3. Deviatoric projection}: $\tau^{\mathrm{ML}}_{ij} \leftarrow \tau^{\mathrm{raw}}_{ij} - \frac{1}{3}\tau^{\mathrm{raw}}_{kk}\,\delta_{ij}$
\State \textbf{4. Stress split} (at each lattice site):
\State \quad $\Pi \leftarrow -\tau^{\mathrm{ML}}_{ij} S_{ij}$; \quad $|\!S\!|^2 \leftarrow 2 S_{ij} S_{ij}$
\State \quad $\nu_t^{\mathrm{eff}} \leftarrow \Pi / |\!S\!|^2$; \quad $\nu_e \leftarrow \nu_0 + \nu_t^{\mathrm{eff}}$
\State \quad $s_\nu \leftarrow 1/(\frac{1}{2} + 3\nu_e)$
\State \quad $\tau^{\mathrm{diss}}_{ij} \leftarrow -2\nu_t^{\mathrm{eff}} S_{ij}$; \quad $\tau^{\mathrm{res}}_{ij} \leftarrow \tau^{\mathrm{ML}}_{ij} - \tau^{\mathrm{diss}}_{ij}$
\State \textbf{5. Coupling}:
\State \quad $\mathbf{F}^{\mathrm{res}} \leftarrow -\nabla\cdot\boldsymbol{\tau}^{\mathrm{res}}$ (centered differences, periodic BCs)
\State \quad $\mathbf{F}_{\mathrm{total}} \leftarrow \mathbf{F}_{\mathrm{ext}} + \mathbf{F}^{\mathrm{res}}$
\State \quad Apply MRT collision (Eq.~\ref{eq:mrt_ml_les}) with $\boldsymbol{\Lambda}^{\mathrm{ML}}(s_\nu)$ and $\mathbf{m}^{F_{\mathrm{total}}}$
\end{algorithmic}
\end{algorithm}

Table~\ref{tab:runtime_mlups} reports end-to-end timing for a representative rollout to $t/t_0=40$ for a dynamic Smagorinsky LES baseline and the ONNX-coupled ML--LES variant, using the same executable and hardware configuration (NVIDIA A100 80~GB GPU; ONNX Runtime GPU v1.16.3; CUDA v12.2). The measured throughput differs by $\approx 0.6\%$.

\begin{table}[t]
\centering
\caption{End-to-end performance comparison for a representative rollout to $t/t_0=40$.}
\label{tab:runtime_mlups}
\begin{tabular}{lccc}
\toprule
Closure & Elapsed time (s) & MLUP/s & Relative diff. (MLUP/s) \\
\midrule
Dynamic Smagorinsky LES & 2929.319 & 71.592 & -- \\
ML--LES (ONNX inference) & 2947.285 & 71.155 & $-0.61\%$ \\
\bottomrule
\end{tabular}
\end{table}

The wall-clock timing in Table~\ref{tab:runtime_mlups} includes both feature computation and inference. In our tested configuration, the ML-coupled rollout achieves throughput comparable to the dynamic Smagorinsky baseline.

\section*{Declaration of competing interest}
The authors declare no competing interests.

\section*{Data availability}
The data that support the findings of this study are available from the corresponding author upon reasonable request.

\section*{Acknowledgments}
The authors wish to acknowledge the support of the National Center for HPC, Big Data and Quantum Computing, Project CN\_00000013---CUP E83C22003230001, Mission 4 Component 2 Investment 1.4, funded by the European Union---NextGenerationEU; the support of Project PRIN 2022F422R2---CUP E53D23003210006, financed by the European Union---Next Generation EU; the support of Project PRIN PNRR P202298P25---CUP E53D23016990001, financed by the European Union---Next Generation EU; and the support of Project ECS 0000024 Rome Technopole---CUP B83C22002820006, NRP Mission 4 Component 2 Investment 1.5, funded by the European Union---NextGenerationEU.

We used KI-TURB 3D~\cite{ki_turb_3d} for literature and code search during manuscript preparation.

\clearpage
\bibliographystyle{elsarticle-num}
\bibliography{references.bib}

\end{document}